# Comparative investigation of third and fifth harmonic generation in atomic and molecular gases driven by mid-infrared ultrafast laser pulses

Jielei Ni [1,2], Jinping Yao [1,2], Bin Zeng [1,2], Wei Chu [1,2], Guihua Li [1,2], Haisu Zhang [1,2], Chenrui Jing [1,2], See Leang Chin [3], Ya Cheng [1,†], Zhizhan Xu [1,‡]

[1] *State Key Laboratory of High Field Laser Physics, Shanghai Institute of Optics and Fine Mechanics, Chinese Academy of Sciences, P.O. Box 800-211, Shanghai 201800, China*

[2.] *Graduate School of the Chinese Academics of Sciences, Beijing 100039, China*

[3.] *Department of Physics, Engineering Physics and Optics & Center for Optics, Photonics and Laser (COPL), Laval University, Laval, Quebec, Canada G1K 7P4*

[†]*Email: ya.cheng@siom.ac.cn*
[‡]*Email: zzxu@mail.shcnc.ac.cn*

**Abstract:**

We report on the comparative experimental investigation on third and fifth harmonic generation (THG and FHG) in atomic and molecular gases driven by mid-infrared ultrafast laser pulses at a wavelength of ~1500 nm. We observe that the conversion efficiencies of both the THG and FHG processes saturate at similar peak intensities close to ~$1.5\times10^{14}$ W/cm$^2$ for argon, nitrogen, and air, whose ionization potentials are close to each other. Near the saturation intensity, the ratio of yields of the FHG and THG reaches ~$10^{-1}$ for all the gases. Our results show that high-order Kerr effects do exist, whereas they are insufficient to balance the Kerr self-focusing without the assistance of plasma generation.

**PACS numbers:** 42.65.Re, 42.68.Ay

## I. INTRODUCTION

Recently, a remarkable finding reported by V. Loriot et al [1,2] has motivated studies on the role of high-order Kerr effect (HOKE) in filamentation. The group found a negative HOKE at the laser intensity above $2.6 \times 10^{13}$ W/cm$^2$ and proposed that HOKE should be taken into consideration in ultrashort laser filamentation. Previously, the physical mechanism of filamentation is interpreted as the dynamical balance of plasma defocusing and Kerr self-focusing which is defined as $n_2 I$, with $I$ the laser intensity, and $n_2$ the second-order Kerr refractive index coefficient related to the nonlinear susceptibility $\chi^{(3)}$ [3,4]. Such mechanism is referred to as the standard model. The new finding from V. Loriot et al. [1,2] extends the Kerr effect to higher-order terms ($n_2 I + n_4 I^2 + n_6 I^3 + n_8 I^4 + ...$), which is known as the high-order Kerr model. Further theoretical simulations based on high-order Kerr model show that the HOKE can become the dominant defocusing mechanism instead of plasma defocusing [5–8]. From this point of view, many new features are revealed in the process of filamentation, such as lower maximum intensity [5], longer filament length [5,6], higher power ratio of 5$^{th}$ harmonic efficiency to 3$^{rd}$ harmonic efficiency (~10$^{-1}$) [7], and ionization-free filamentaion [5]. High-order Kerr refractive indices used in these theoretical investigations are based on the values measured at 800 nm wavelength by V. Loriot et al. [1,2]. However, subsequent experiments [9–11] which measure either the electron density [9] inside a filament or the conical emission rings produced by filamentation [10] and theoretical simulations [12] show that the obtained data agree better with the standard model which completely neglects the HOKE terms. Furthermore, a direct measurement of the nonlinear optical response by J. K. Wahlstrand et al. [13] shows that neither negative nonlinear refractive index nor saturation can be observed, in contrast with the previous measurements. On the other hand, experimental evidences of existence of the HOKE are reported in Refs. [14,15] recently. Thus, until now a common conclusion has not yet been reached on this dispute.

In 2010, M. Kolosik et al. proposed that the validity of the HOKE model can be experimentally tested by measuring and comparing the conversion efficiencies of $5^{th}$ harmonic generation (FHG) and $3^{rd}$ harmonic generation (THG) processes in gas media [7]. If the high-order Kerr model is correct, the power ratio of $5^{th}$ harmonic (FH) to $3^{rd}$ harmonic (TH) will saturate at a relatively high value of $\sim 10^{-1}$; whereas if the standard model is correct, the ratio remains very small ($10^{-4}$). Indeed, a recent experiment [11] performed at 2.2 μm shows that the FH/TH power ratio reaches $10^{-4}$, which supports the standard model. However, this experiment was carried out in ambient air that propagation effects (such as Guoy phase shift) cannot be completely ruled out. In addition, the measurement was only done at one specific pump laser intensity, thus information on how the THG and FHG evolve with the increasing pump intensity is lacking. In this paper, we carry out the comparative experimental investigation on THG and FHG in atomic and molecular gases driven by mid-infrared ultrafast laser pulses at a wavelength of ~1500 nm. In particular, we chose to use low gas pressure and short interaction length in our experiment in order to minimize the influence from propagation effects as much as possible. Moreover, we compare our results for three kinds of gases, namely, argon, nitrogen, and air. These gases have a similar ionization potential whilst their values of $n_4$ reported in Ref. [1,2] are different. We systematically investigate the conversion efficiencies of THG and FHG as a function of pump intensity in a broad intensity range from $3\times10^{13}$ W/cm$^2$ to $3\times10^{14}$ W/cm$^2$. At the highest pump intensity, we clearly observe the saturation of both the THG and FHG processes.

## II. EXPERIMENT

The experimental setup is shown in Fig. 1. The experiment is performed with ~50 fs (FWHM), 1500 nm laser pulses generated by an optical parametric amplifier (OPA) (HE-TOPAS, Light Conversion, Inc.) pumped by a commercial Ti:sapphire laser system (Legend Elite-Duo, Coherent, Inc.). The Ti:sapphire laser, operated at a repetition rate of 1 kHz, provides ~40 fs (FWHM) laser pulses with a central wavelength at ~795 nm and single pulse energy of 6.4 mJ. At the 1500 nm wavelength, the maximum pulse energy that the OPA can offer is ~1.3 mJ, and the pulse duration is ~48 fs. After being separated from the idler using two reflection mirrors with high reflectivity at 1500 nm (M1, M2), the infrared beam is focused into a 2-mm-long gas cell by a fused silica lens (L1) with a focal length of 31.4 cm. The focal length is measured at 1500 nm wavelength taking into account the 3-mm-thick window of the vacuum chamber. Back pressure inside the vacuum chamber is ~0.1 Pa. The gas cell mounted inside the vacuum chamber is filled with gas for filamentation. Argon, air, and nitrogen are used in this experiment with a gas pressure of 50 mbar.

Harmonic beams generated in the filament are collimated by another fused silica lens (L2) with a focal length of ~40 cm, and then the beam diameters are reduced by a telescope system consisted of a plano-convex lens with the focal length of 50 cm (L3) and a plano-concave lens with the focal length of -20 cm (L4). Finally, the harmonics are impinging onto the adjustable entrance slit of an imaging grating spectrometer (Shamrock 303i, Andor). The slit size is comparable to the diameter of harmonic beams. Input laser intensity is adjusted by simply changing the incident angle of a thin glass plate before L1. Attenuator for TH or FH is inserted between the lenses L2 and L3.

Nonlinear response from the attenuator can be excluded by their constant attenuation factor measured over a wide range of input intensity. Harmonic powers are measured by integrating the recorded spectra over the interested spectral range. Calibration is achieved with the 800 nm laser pulses from the amplifier, and the variation of diffraction efficiency of the imaging grating as well as the quantum efficiency of the CCD with wavelength are taken into account.

## III. RESULTS AND DISCUSSION

Figures 2(a) and (b) show the intensity dependence of the harmonic conversion efficiency and the power ratio of FH to TH in argon, respectively. As illustrated in Fig. 2(a), the conversion efficiencies first increase following the third-power law dependence for THG and the fifth-power law dependence for FHG, indicating that the generation of the TH and FH can be well described by perturbative theory. The curves then deviate from these perturbative power laws and both saturate at the laser intensity of $\sim 1.5 \times 10^{14}$ W/cm$^2$. The FH/TH power ratio reaches the order of $\sim 10^{-1}$ around the saturation intensity, as shown in Fig. 2(b). Similar results are obtained in air and nitrogen with a saturation intensity of about $1.5 \times 10^{14}$ W/cm$^2$, and the FH/TH power ratio again reaches $\sim 10^{-1}$, as demonstrated in Figs. 2(c)-(f).

In order to interpret the saturation behavior in both the THG and FHG measurements, two possible mechanisms are considered. The first mechanism is the depletion of ground state, which occurs when the peak intensity is high enough that all the neutral atoms near the focus are ionized [16]. For this mechanism to take effect, ionization probability should be close to unity at the saturation intensity measured in our experiment (i. e., $\sim 1.5 \times 10^{14}$ W/cm$^2$). However, a calculation using Ammosov-Delone-Krainov model shows that, in our experiment, the ionization probability is only ~3% at this saturation intensity, which is far less than unity. Thus the saturation of harmonic generation cannot be attributed to the depletion of the ground state. An alternative mechanism is the intensity clamping effect, which sets an upper limit to the laser intensity inside the filament due to the dynamic balance between the Kerr self-focusing and either plasma defocusing or the recently proposed

HOKE, resulting in a stabilized harmonic intensity. Since the intensity used in this experiment reaches the critical intensity for filamentation, we conclude that the intensity clamping effect should be responsible for the saturation in harmonic generation. It is noteworthy that although we have attempted to minimize the propagation effects in our experiment using the thin gas sample, however, as there are two holes drilled at the two ends of the gas cell, leakage of gas can occur. Thus, at high pump intensities, intensity clamping can still occur which gives rise to the saturated conversion efficiencies of both THG and FHG processes. Below, we will discuss whether the HOKE or the plasma defocusing plays the dominate role in balancing the Kerr focusing effect.

It is noticeable that in our experiment, the FH/TH power ratios reach $\sim 10^{-1}$ when saturation occurs as shown in Figs. 2(b), (d), and (f). As pointed out by M. Kolesik et al. [6], in the standard model the $5^{th}$ harmonic arises via a cascade process (e. g., $5\omega = 3\omega + \omega + \omega$), leading to a low power ratio of $5^{th}/3^{rd}$ harmonic. However, in the high-order Kerr model, $5^{th}$ harmonic can be generated directly (e. g., $5\omega = \omega + \omega + \omega + \omega + \omega$) from the fifth-order nonlinear susceptibility, resulting in a FH/TH power ratio as high as $10^{-1}$. To further identify whether the FHG is a cascaded or direct process, we recall that, in the perturbative limit, the nonlinear polarization scales as $P^{(n)} = N\chi^{(n)} E_1 E_2 \ldots E_n$, where N is the atomic density. Thus, in the high-order Kerr model, the $5^{th}$ harmonic signal is generated by direct process and scales as $S \propto N^2$. On the other hand, in the standard model, the scaling law of FHG yield on gas pressure should be modified because now FHG is a cascaded process, therefore we shall have $S \propto N^4$. As clearly shown in Fig. 3, the harmonics efficiency increases nearly quadratically with pressure in our experiment, which suggests that the FH is generated via the direct process. In this case, using the equation (6) in Ref [15] and neglecting the phase mismatch for both THG and FHG due to the short interaction length, we have $\frac{P_5}{P_3} = \left|\frac{n_4}{n_2}\right|^2 \frac{3}{5} I^2 = \frac{3}{5}\left|\frac{\Delta n_4}{\Delta n_2}\right|^2$, where $\Delta n_4 = n_4 I^2$ and $\Delta n_2 = n_2 I$. Based on the experimental data in Figs. 2(b, d, f), the ratio of the fourth-order Kerr refractive index to the second-order Kerr refraction index $\left|\Delta n_4/\Delta n_2\right|$ can be obtained. At the intensity of $1.5 \times 10^{14}$ W/cm$^2$, the ratio $\left|\Delta n_4/\Delta n_2\right|$ is evaluated to be 0.49 for argon, 0.48 for air, and 0.57 for nitrogen. These results suggest that the fifth-order nonlinearities cannot be completely neglected in the FHG process until the ionization takes place. However, even if we assume that the fourth-order Kerr refractive index coefficient is negative (the sign of the fourth-order Kerr coefficient cannot be determined in our experiment), the defocusing induced by the fourth-order Kerr effect are still not large enough to balance the second-order-Kerr-effect induced self-focusing. That is to say, in femtosecond laser filamentation, plasma defocusing is still necessary for arresting the self-focusing collapse. This may explain why the plasma generation can always be observed in ultrafast filamentation experiments.

## IV. CONCLUSION

To summarize, we have systematically investigated the intensity dependence of THG

and FHG conversion efficiency in argon, nitrogen, and air. We show that the TH and FH signals increase nicely with the pump intensity following the third and fifth power laws until the plasma defocusing takes place at $\sim 1.5\times 10^{14}$ W/cm$^2$. The power ratio of $5^{th}/3^{rd}$ harmonic saturates at $\sim 10^{-1}$ for all the three gas samples. The measured harmonics efficiencies increase with the squared values of the gas pressures, indicating that the FHG is a direct process instead of a cascaded wave-mixing process. Based on this finding, the fourth-order Kerr refractive index coefficient is evaluated by the measured ratio of the $5^{th}$ to the $3^{rd}$ harmonic, which is around the half of the second-order Kerr refractive index coefficient. Thus, the HOKE is still insufficient to arrest Kerr self-focusing collapse without the assistance from the plasma formation in femtosecond filamentation.

## ACKNOLEDGMENTS

We are grateful to M. Kolesik for the stimulating discussions. This research is financially supported by National Basic Research Program of China (2011CB808100), and National Natural Science Foundation of China (Grant Nos. 10974213, 60825406, 11174156). SLC acknowledges the support from the Canada Research Chairs program.

**Captions:**

Fig. 1 (Color online) Schematic of the experimental setup.

Fig. 2 (Color online) Experimentally measured intensity dependence of harmonic conversion efficiency (a, c, e) and power ratio of $5^{th}$ and $3^{rd}$ harmonic (b, d, f) in argon (a, b), air (c, d) and nitrogen (e, f).

Fig. 3 (Color online) Experimentally measured pressure dependence of harmonic conversion efficiency in argon.

Fig. 1

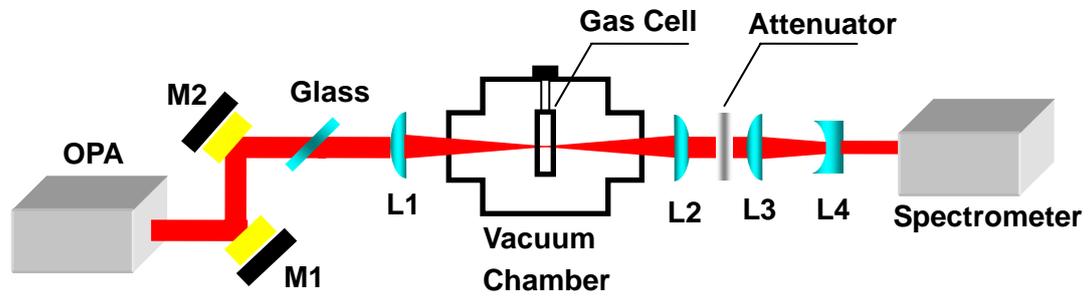

Fig. 2

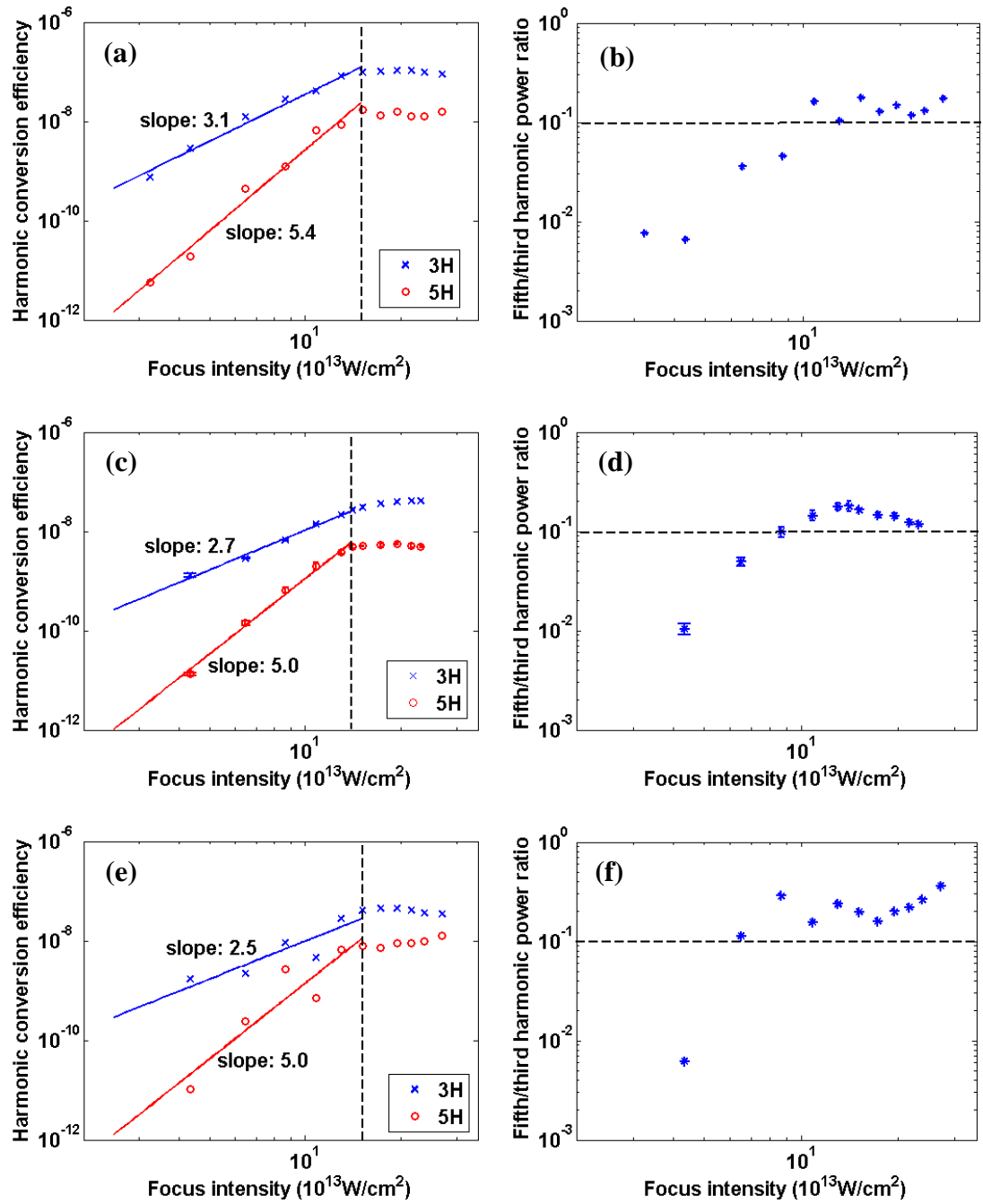

Fig. 3

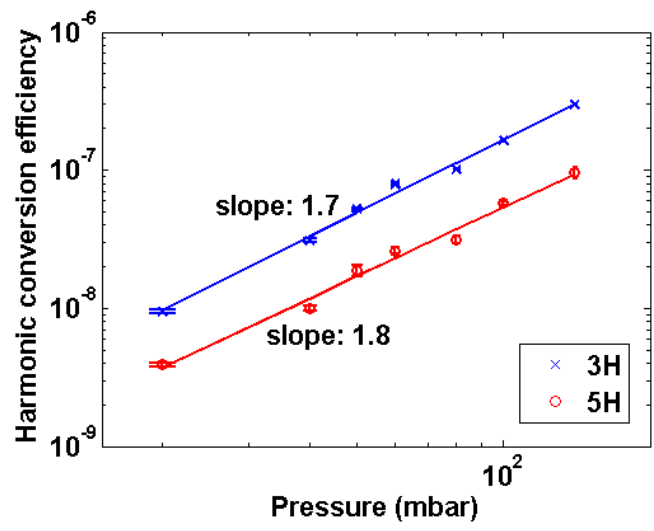